\documentclass[12pt]{article}
\newcommand{\Bbb}[1]{\mathbf{#1}}
\usepackage[dvips]{graphicx}
\usepackage{psfrag}
\makeatletter
 
 \@addtoreset{equation}{section}
\makeatother 
\newcommand{\sumtwo}[2]{\mathop{\sum_{#1}}_{#2}}
\newcommand{\limtwo}[2]{\mathop{\lim_{#1}}_{#2}}
\newcommand{\calP}{\mathcal{P}}
\newcommand{\calK}{\mathcal{K}}
\newcommand{\calB}{\mathcal{B}}
\newcommand{\calS}{\mathcal{S}}
\newcommand{\calD}{\mathcal{D}}
\newcommand{\up}{\uparrow}
\newcommand{\dn}{\downarrow}
\newcommand{\bs}{\backslash}

\newcommand{\sgn}{{\mathbf{sgn}}}
\newcommand{\La}{\Lambda}
\newcommand{\Do}{D_\mathrm{o}}
\newcommand{\De}{D_\mathrm{e}}
\newcommand{\rme}{\mathrm{e}}
\newcommand{\rmi}{\mathrm{i}}

\newcommand{\Nh}{{N_\mathrm{h}}}
\newcommand{\Np}{{N_\mathrm{p}}}
\newcommand{\Nhd}{{N_\mathrm{h}^d}}
\newcommand{\Nhp}{{N_\mathrm{h}^p}}
\newcommand{\vecS}{\mbox{\boldmath $S$}}
\newcommand{\PhiG}{\Phi_{\mathrm{p}}}
\newcommand{\Tc}{T_\mathrm{c}}
\newcommand{\bsigma}{\mbox{\boldmath $\sigma$}}
\newcommand{\btau}{\mbox{\boldmath $\tau$}}
\newcommand{\HZRS}{\mathbf{H}_\mathrm{ZRS}^\Nh}
\newcommand{\ved}{\varepsilon_d}
\newcommand{\vezero}{\varepsilon_0}
\newcommand{\tJmodel}{{$t$-$J$ model}}
\newcommand{\dxs}{d_{x,\sigma}}
\newcommand{\dxsd}{d_{x,\sigma}^\dagger}
\newcommand{\pus}{p_{u,\sigma}}
\newcommand{\pusd}{p_{u,\sigma}^\dagger}
\newcommand{\aks}{\hat{a}_{k,\sigma}}
\newcommand{\axs}{{a}_{x,\sigma}}
\newcommand{\zrxd}{{\psi}_{x}^\dagger}

\newcommand{\tphiz}{\tilde{\phi}_z}
\newcommand{\expec}[1]{{\left< #1 \right>}}
\newcommand{\expPhiG}[1]{{\left<\PhiG, #1 \PhiG\right>}}
\newcommand{\eqref}[1]{(\ref{#1})}
\begin{document}
%\baselineskip=1.5\baselineskip
%   title   
\begin{center}
\textbf{\large 
 $d$-Wave Pairing State in Terms of the Zhang-Rice Singlets
}\bigskip\\
Akinori Tanaka%
\footnote{
 akinori@ariake-nct.ac.jp
}\bigskip\\
\textit{Department of General Education, 
          Ariake National College of Technology,
        Omuta, Fukuoka~836-8585, Japan}\bigskip\\
(April 25, 2007)
\end{center}
\vspace*{4cm}
\begin{abstract}
%\baselineskip=1.5\baselineskip
In cuprate superconductors doping is believed to create holes on the O-sites,
which couple antiferromagnetically with holes on the Cu-sites
to form the so-called Zhang-Rice singlets.
Here we study a $d$-wave pairing state based on the Zhang-Rice singlet
states. Upper and lower bounds of an off-diagonal long-range order parameter 
with $d$-wave symmetry for this state are estimated.
We also introduce a concrete model with on-site Coulomb repulsion and
kinds of antiferromagnetic interactions whose ground state is
this $d$-wave pairing state.
\end{abstract}
\vfill
\if0
{Short title}:
 \textit{$d$-Wave Pairing State in Terms of the Zhang-Rice Singlets}
\fi
\newpage
%%%%%%%%%%%%%%%%%%%%%%%%%%%%%%%%%%%
%%%%%%%%%%%%%%%%%%%%%%%%%%%%%%%%%%%
%%%   main text 
%%%%%%%%%%%%%%%%%%%%%%%%%%%%%%%%%%%
%%%%%%%%%%%%%%%%%%%%%%%%%%%%%%%%%%%
\section{Introduction}
The mechanism of high-$\Tc$ cuprate superconductivity has been attracting much
interest since it is discovered in 1986~\cite{Bednorz}.
In cuprate superconductors, electrons (or holes) in the CuO$_2$ planes
play major roles, and 
the importance of the Coulomb repulsion at the Cu-sites is emphasized
from the beginning~\cite{Anderson,Emery87,Hirsch87}. 
However, theoretical understanding of its effects on the superconductivity is
still limited and is being a challenging problem in condensed matter physics.

Most theories which start with viewing cuprate superconductors as doped
Mott insulators are based on the so-called Zhang-Rice singlet states~\cite{ZhangRice88}.
In the undoped case, where there is one hole per Cu-site in CuO$_2$ planes, 
the cuprates exhibit insulating antiferromagnetism due to 
the strong Coulomb repulsion at the Cu-sites.
When the system is doped, additional holes are created on the O-sites.
Because of a superexchange antiferromagnetic interaction,
each of the holes occupies a quasi-localized state on the four nearest
neighbour O-sites around a Cu-site, forming a local spin-singlet with the hole on
the central Cu-site.      
This singlet is now referred to as Zhang-Rice singlet. 
The Zhang-Rice singlets become charge carriers moving through the
CuO$_2$ plane and condense into a superconducting state. 

This scenario is usually examined by using 
the \tJmodel~which is a single-band effective
Hamiltonian with antiferromagnetic interactions between
nearest neighbour holes on
the Cu-sites~\cite{ZhangRice88}.
Despite its simple form, however, it is a formidably difficult task to
rigorously analyze the \tJmodel, and whether the model really describes
the cuprate superconductivity has not yet been clarified.
In the current situation, we think that a rigorous establishment of
occurrence of a superconducting state based on the Zhang-Rice singlets 
in a model with the Coulomb repulsion and
antiferromagnetic interactions, even if it is apart from the \tJmodel,
certainly gives us an important step toward understanding of the
cuprates superconductivity.

In this paper, we study a simple $d$-wave pairing state expanded in terms of
the Zhang-Rice singlet states. 
It is shown that the pairing state is regarded as a condensed state of
the Zhang-Rice singlets in the background of a resonating-valence-bond
state consisting of holes at the Cu-sites.
We estimate an upper bound on an off-diagonal long-range order (ODLRO) parameter with
$d$-wave symmetry for the pairing state as a function of doping
concentration $0\le\delta\le1$. It is found that an upper bound has a dome structure
with a maximum at $\delta=0.5$ and  becomes zero at $\delta=0,1$.   
We also estimate a lower bound on the ODLRO parameter and show 
that ODLRO exists for sufficiently large doping concentrations.
We then introduce a model with  on-site repulsion and kinds of
antiferromagnetic interactions, 
and show that the pairing state is a ground state of this model.
A related model with infinitely large on-site repulsion at the Cu-site 
is analyzed in Ref.~\cite{Tanaka04}. 
This model, however, has following disadvantages:
the Hamiltonian does not have spin rotational symmetry, and
its exact pairing ground state has less relevance to the Zhang-Rice singlets. 
Although the present model has still somewhat artificial aspects, it is for the
first time that the pairing state with $d$-wave symmetry 
which is written explicitly in terms of the Zhang-Rice singlet states 
is realized as a ground state of the concrete Hamiltonian.

This paper is organized as follows.
In the next section we prepare some notation and give a definition of 
the Zhang-Rice singlet states.
In section~\ref{sc:Pairing State}, we introduce a two-electron state with
$d$-wave symmetry, and,
on the basis of the Zhang-Rice singlet states, we construct a pairing
state in which many electrons condense into this two-electron state.
In section~\ref{sc:Order Parameter}, we discuss
an expectation value of an order parameter with $d$-wave
symmetry for the pairing state.
An upper bound for the order parameter is obtained in this section and 
a lower bound, whose estimation needs somewhat technical calculations, is
obtained in section~\ref{sc:Lower Bound}.
In section~\ref{sc:Hamiltonian} we introduce a Hamiltonian whose ground
state is the pairing state which we construct.
In the final section, a summary and some remarks are given.
In Appendix~\ref{a:non-vanishing} we show that the pairing state 
is non-vanishing.
%%%%%%%%%%%%%%%%%%%%%%%%%%%%%%%%%%%
\section{Zhang-Rice singlet states}
\label{sc:Zhang-Rice}
\begin{figure}[t]
\begin{center}
 \includegraphics[width=.3\textwidth]{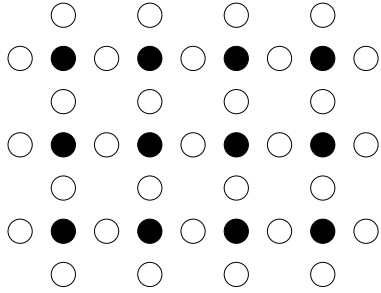}
\end{center}
\caption{
The lattice structure.
The solid and open circles indicate the Cu- and O-sites, respectively.
}
\label{fig:lattice}
\end{figure}

We start with the definition of a lattice.
With even integers $L_1$ and $L_2$, 
let
\begin{equation}
 D=\left([1,L_1]\times[1,L_2]\right)\cap \Bbb{Z}^2, 
\end{equation}
which represents
a collection of the Cu-sites. 
Let $\delta^1=(1,0)$ and $\delta^2=(0,1)$.
We define
\begin{equation}
 P=\{u~|~u=x+\delta^l/2,~l=1,2,~x\in D\}, 
\end{equation}
which is the collection of the mid-points of 
the nearest neighbour bonds in $D$ and
corresponds to the O-sites. 
Then we consider the lattice $\La = D\cup P$, which
mimics the CuO$_2$ plane. (See Fig.~\ref{fig:lattice}.)
For a technical reason we impose periodic boundary conditions on $\La$.
For later use, we introduce further the following sublattices of $D$:
\begin{eqnarray}
 \Do &=&\{x~|~\mbox{$x=(x_1,x_2)\in D$ with $x_1+x_2$ being odd}\},\\
 \De &=&\{x~|~\mbox{$x=(x_1,x_2)\in D$ with $x_1+x_2$ being even}\}.
\end{eqnarray}
%%%%%%%%%%%%%%%%%%%%%%%%%%%%%%%%%%%%%%%%%%%%%%%%%%%%%%%%%%%%%%%%%%%%%%%%%%%%%%%
%%%%%%%%%%%%%%%%%%%%%%%%%%%%%%%%%%%%%%%%%%%%%%%%%%%%%%%%%%%%%%%%%%%%%%%%%%%%%%%
%%%%%%%%%%%%%%%%%%%%%%%%%%%%%%%%%%%%%%%%%%%%%%%%%%%%%%%%%%%%%%%%%%%%%%%%%%%%%%%
%%%%%%%%%%%%%%%%%%%%%%%%%%%%%%%%%%%%%%%%%%%%%%%%%%%%%%%%%%%%%%%%%%%%%%%%%%%%%%%

Next we introduce fermion operators 
which annihilate or create \textit{holes} 
with spin $\sigma=\up,\dn$ at sites in $\La$.
Any states with the number $\Nh$ of holes can be constructed by
operating these operators on a state $\Phi_0$ with no holes on $\La$.
By $\dxs(\dxsd)$ and $\pus(\pusd)$,  
we denote the annihilation(creation) operators of holes
at $x\in D$ and $u\in P$, respectively.  
As mentioned in section 1, 
each hole additionally induced in a CuO$_2$ plane with 1 hole per Cu
is considered to localize well at the four nearest O-sites of a Cu-site
because of the antiferromagnetic superexchange interactions between 
Cu- and O-sites.
To describe this localized state on the O-sites
we introduce the following operators for each $x\in D$~\cite{comment1}:
\begin{equation}
 f_{x,\sigma}=\frac{1}{2}\sum_{u\in P;|u-x|=1/2} \pus.
 \label{eq:fxs}
\end{equation}
As is easily seen, the annihilation operator $f_{x,\sigma}$ 
and the creation operator $f_{x^\prime}^\dagger$ defined by
\eqref{eq:fxs} do not anticommute when $|x-x^\prime|=1$,
implying that the single-electron states corresponding to \eqref{eq:fxs}
are not orthogonal.
To avoid technical complexities arising from this fact,
we consider corresponding Wannier states.
To do so, we introduce the fermion operator 
$f_{\sigma}^{1}=(1/\sqrt{D})\sum_{x\in D}
\rme^{\rmi\pi\delta^1\cdot x}p_{x+\delta^1/2,\sigma}$
and the reciprocal lattice
\begin{equation}
 \calK=\left\{(\frac{2\pi}{L_1}n_1,\frac{2\pi}{L_2}n_2)~|~
                         ~n_l\in \Bbb{Z},~-L_l/2<n_l\le L_l/2
                          ~\mbox{with $l=1,2$} \right\},
\end{equation}
and then define 
$\hat{f}_{k,\sigma}=(1/\sqrt{|D|})\sum_{x\in D} f_{x,\sigma} 
\rme^{-\rmi k\cdot x}$
for $k\in \calK\bs\{(\pi,\pi)\}$ and 
$\hat{f}_{(\pi,\pi),\sigma}=f_{\sigma}^1$.
We normalize the $\hat{f}$-operators as 
$\aks=\hat{f}_{k,\sigma}/\parallel f_{k} \parallel$,
where the normalization factors are given by
\begin{equation}
 \parallel f_{k} \parallel
    =\left\{ \begin{array}{@{\,}ll}
                1 & \mbox{if $k=(\pi,\pi)$,}\\
	      \sqrt{1+\frac{1}{2}(\cos k_1+ \cos k_2)} & \mbox{otherwise.}
             \end{array}
      \right.
\end{equation}
The fermion operators corresponding to the Wannier states are
defined by
\begin{equation}
  \axs = \frac{1}{\sqrt{|D|}}\sum_{k\in \calK} \aks \rme^{\rmi k\cdot x}.
\end{equation}
The $a$-operators defined as above approximate the $f$-operators well, and satisfy 
the canonical fermion anticommutation relations 
$\{a_{x,\sigma}^\dagger,a_{y,\tau}^\dagger\}=
\{a_{x,\sigma},a_{y,\tau}\}=0$ and
$\{a_{x,\sigma}^\dagger,a_{y,\tau}\}=\delta_{\sigma,\tau}\delta_{x,y}$
for $\sigma,\tau=\up,\dn$ and $x,y\in D$.
In the rest of this paper, we consider the Zhang-Rice singlets 
by using the $a$-operators, instead of the $f$-operators.
%%%%%%%%%%%%%%%%%%%%%%%%%%%%%%%%%%%%%%%%%%%%%%%%%%%%%%%%%%%%%%%%%%%%%%%%%%%%%%%
%%%%%%%%%%%%%%%%%%%%%%%%%%%%%%%%%%%%%%%%%%%%%%%%%%%%%%%%%%%%%%%%%%%%%%%%%%%%%%%
\begin{figure}[t]
\label{fig:state}
\begin{center}
 \begin{tabular}{lcl}
(a) & \hspace*{.1\textwidth}& (b) \\
 \includegraphics[width=.3\textwidth]{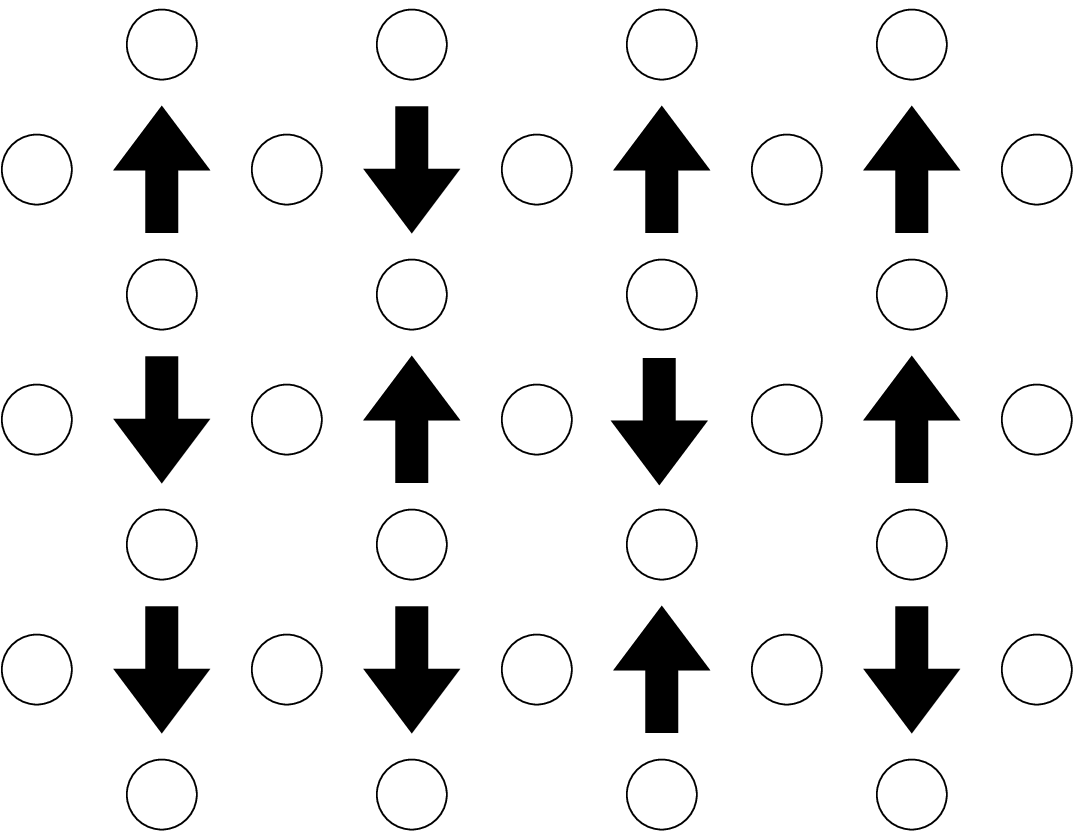} &&
 \includegraphics[width=.3\textwidth]{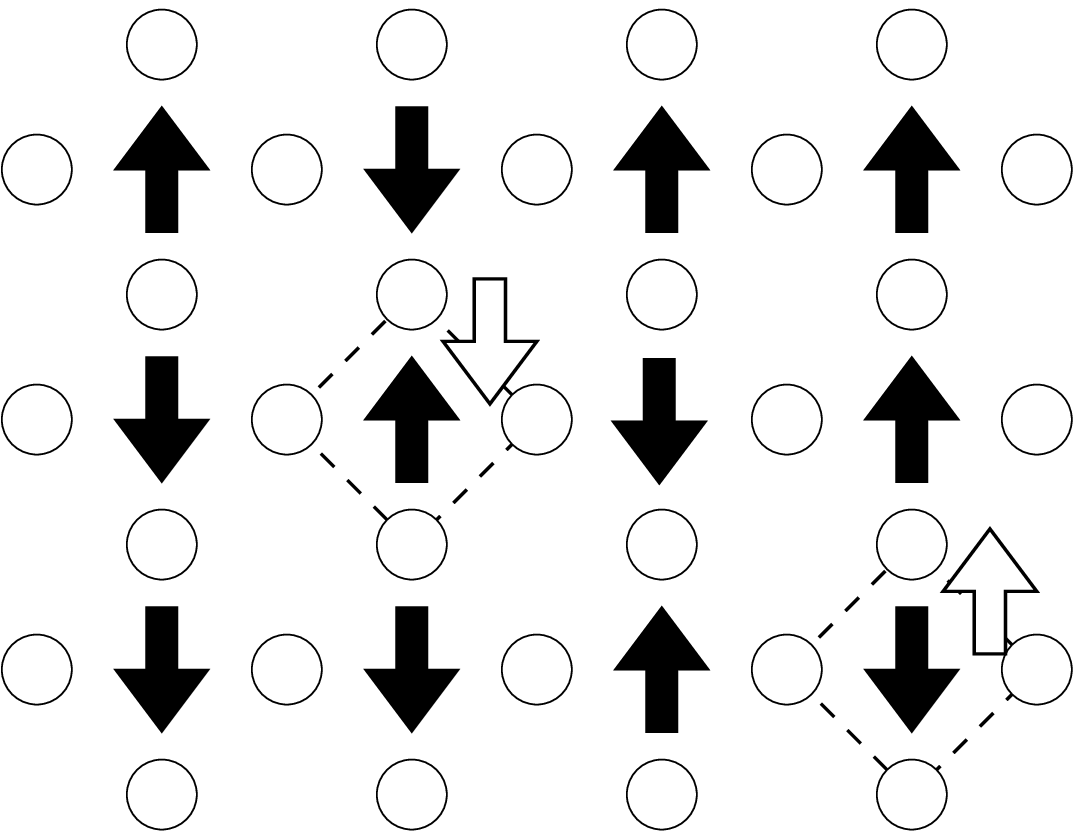} 
\end{tabular}
\end{center}
\caption{
The solid and open arrows indicate spins of the holes on the Cu- and
 O-sites, respectively.
(a) In the case of $\Nh=|D|$, every Cu-site is occupied by one
 hole. 
(b) When $\Nh$ is greater than $|D|$, every Cu-site remains to be
 occupied by one hole, and additional holes are created on the O-sites.
 Each hole on the O-sites occupies a quasi-localized state, 
 which is indicated by dot lines, and couples to the hole at 
 the central Cu-site to form the Zhang-Rice singlet.   
}
\end{figure}
%%%%%%%%%%%%%%%%%%%%%%%%%%%%%%%%%%%%%%%%%%%%%%%%%%%%%%%%%%%%%%%%%%%%%%%%%%%%%%%

The Zhang-Rice singlet around a Cu-site $x$ is formed by holes occupying
$a_{x,\sigma}^\dagger$ and $d_{x,\tau}^\dagger$. This singlet is represented by 
the two-hole creation operator 
\begin{equation}
 \zrxd =  d_{x,\up}^\dagger a_{x,\dn}^\dagger
            +a_{x,\up}^\dagger d_{x,\dn}^\dagger.  
\end{equation}

We assume that,
in the case where the hole number is $|D|$, each hole occupies a Cu-site.
Any $|D|$-hole state is then  
expressed by a linear combination of 
$\prod_{x\in D}d_{x,\sigma_x}^\dagger\Phi_0$ with $\sigma_x=\up,\dn$ (Fig.~\ref{fig:state}(a)).  
We furthermore assume that $N$ holes added in this state form Zhang-Rice
singlets. 
Then a $(|D|+N)$-hole state with 
$0< N \le |D|$ is written by using
a set of states
\begin{equation}
 \left(\prod_{x\in A}d_{x,\sigma_x}^\dagger \right)
 \left(\prod_{y\in D\bs A}\psi_{y}^\dagger \right)\Phi_0, 
\label{eq:ZRS-basis0}
\end{equation}
where $A$ is a subset of $D$ with $|A|=|D|-N$ and its compliment $D\bs A$ 
is a collection of sites where the Zhang-Rice singlets are formed (Fig.~\ref{fig:state}(b)). 
Noting the relation
\begin{equation}
 d_{x,\sigma}^\dagger\Phi_0=-\mathbf{sgn}[\sigma]a_{x,-\sigma}\zrxd\Phi_0
\end{equation}
where $\mathbf{sgn}[\sigma]=+$ if $\sigma=\up$ and 
$\mathbf{sgn}[\sigma]=-$ if $\sigma=\dn$, we find that
\eqref{eq:ZRS-basis0} is rewritten as
\begin{equation}
 \left(\prod_{x\in A}a_{x,-\sigma_x} \right)\Psi_0
 = \Psi_{A,\bsigma_{A}},
\label{eq:ZRS-basis}
\end{equation}
with
\begin{equation}
\Psi_0 = \left(\prod_{y\in D}\psi_y^\dagger \right)\Phi_0 
\end{equation} 
up to a sign factor.
Here $\bsigma_A$ is a short hand for a spin configuration
$(\sigma_x)_{x\in A}$.
We write  $\calS_A$  for the collection of spin configurations
$\{(\sigma_x)_{x\in A}~|~\sigma_{x}=\up,\dn,~x\in A\}$. 
It is easy to see that 
$\left< \Psi_{A,\bsigma_A}, \Psi_{B,\btau_B} \right>
= 2^{|D|-|A|}\chi[A=B]\chi[\bsigma_A=\btau_B]$,
where $\chi[\mathrm{event}]=1$ if `event' is true and 0 otherwise.
Thus the collection of states
\begin{equation}
\{\Psi_{A,\bsigma_A}~|~A\subset D,~\bsigma_A\in\calS_A\}
\end{equation} 
is orthogonal.
For $|D|< \Nh \le 2|D|$, 
let $\HZRS$ be the Hilbert space spanned by the basis states 
$\left\{\Psi_{A,\bsigma_A}\right\}$ with $|A|=2|D|-\Nh$.
The Zhang-Rice singlet states are defined to be states in $\HZRS$.    
%%%%%%%%%%%%%%%%%%%%%%%%%%%%%%%%%%%
\section{$d$-Wave Pairing State}
\label{sc:Pairing State}
Assuming that $\Nh$ takes an even number in 
$|D| < \Nh \le 2|D|$, we consider a $d$-wave pairing state 
in the Hilbert space $\HZRS$.
Let us define a pair operator $\zeta$ by 
\begin{equation}
 \zeta=\sum_{k=(k_1,k_2)\in\calK}(\cos k_1-\cos k_2)\hat{a}_{-k,\dn} \hat{a}_{k,\up}.
\end{equation}
Recall that
$\hat{a}_{k,\sigma}=(1/\sqrt{|D|})\sum_{x\in D} a_{x,\sigma}\rme^{-\rmi k\cdot x}$
are the Fourier transforms of $a_{x,\sigma}$.
This operator \textit{creates} an \textit{electron} pair 
with $d$-wave symmetry. 
For the hole number $\Nh=|D|+N$ with a positive even integer $N$, 
one of the simplest $d$-wave pairing states in $\HZRS$ is given by
\begin{equation}
 \PhiG = \left(\zeta\right)^\Np\Psi_0
\label{eq:pairing-state}
\end{equation}
with the number of pairs $\Np=(|D|-N)/2$.
Here we note that  
\begin{equation}
 a_{x,\dn}a_{x,\up}\Psi_0=0
\label{eq:no-double-occupancy}
\end{equation}
since 
$[a_{x,\dn}a_{x,\up},\zrxd]=-d_{x,\up}^\dagger a_{x,\up}
                            -d_{x,\dn}^\dagger a_{x,\dn}$,
so that $\PhiG$ is actually expanded in terms of \eqref{eq:ZRS-basis}.
In Appendix \ref{a:non-vanishing} we show that $\PhiG$ is non-vanishing.

In order to see the real space representations of $\zeta$ 
and $\PhiG$, we define the following operators: 
for $x\in D$ 
\begin{equation}
 b_{x,\sigma}
  =\frac{1}{2}\sum_{y\in D;|x-y|=1}
       a_{y,\sigma}\rme^{\rmi\pi\delta^2\cdot(x-y)},
\end{equation}
and for $x,y\in D$
\begin{equation}
 \phi_{\{x,y\}}^a = \frac{1}{2}\rme^{\rmi\pi \delta^2\cdot(x-y)}
                               (a_{x,\dn}a_{y,\up}+a_{y,\dn}a_{x,\up}).
\end{equation}
The operator $\phi_{\{x,y\}}^a$ corresponds to a two-electron singlet
state formed by electrons on the O-sites
around Cu-sites $x$ and $y$. 
By using these operators we can write $\zeta$ as
\begin{equation}
 \zeta=\sum_{x\in D} a_{x,\dn}b_{x,\up}=\sum_{x\in D} b_{x,\dn}a_{x,\up}
\label{eq:zeta3}
\end{equation} 
or 
\begin{equation}
 \zeta=\sum_{\{x,y\}\in \calB}\phi_{\{x,y\}}^a,
\label{eq:zeta2}
\end{equation}
where 
\begin{equation}
 \calB=\left\{ \{x,y\}~|~x,y\in D,~|x-y|=1 \right\}
\end{equation}
is the collection of bonds in $D$ (we assume that $\{x,y\}=\{y,x\}$).
Let $C({\calB})$ be the collection of subsets $B$ of $\calB$
such that no two elements in $B$ share the same site. 
Substituting \eqref{eq:zeta2} into \eqref{eq:pairing-state}, 
and noting the relation \eqref{eq:no-double-occupancy}, we obtain
\begin{equation}
  \PhiG=\Np!\sum_{B\in C(\calB);|B|=\Np}
             \prod_{\{x,y\}\in B}\phi_{\{x,y\}}^a\Psi_0.         
\end{equation}
Therefore, the pairing state $\PhiG$ is regarded as 
a nearest-neighbour resonating-valence-bond state
(which is a linear combination of products of two-electron singlets)
consisting of electrons on O-sites 
with the background of the fully-filled Zhang-Rice singlets. 

Finally let us see the form of $\PhiG$ in terms of the hole creation
operators.
Let $n_{x,\sigma}^d=d_{x,\sigma}^\dagger d_{x,\sigma}$ and define 
$\calP_{D}=\prod_{x\in D}(1-n_{x,\up}^{d}n_{x,\dn}^d)$, which
is the projection operator onto the space without double occupancies of
holes at the Cu-sites.
By using this projection operator we can rewrite $\Psi_0$ as
\begin{equation}
 \Psi_0 =\frac{1}{|D|!} \calP_{D}
            \left(\sum_{x\in D} \psi_{x}^\dagger \right)^{|D|}\Phi_0
        =\frac{1}{|D|!} \calP_{D}
            \left(\sum_{k\in \calK} 
                  \left(\hat{d}_{k,\up}^\dagger\hat{a}_{-k,\dn}^\dagger
                    + \hat{a}_{k,\up}^\dagger \hat{d}_{-k,\dn}^\dagger\right)
            \right)^{|D|}\Phi_0,
\end{equation}
 where $\hat{d}_{k,\sigma}
         =(1/\sqrt{|D|})\sum_{x\in D} d_{x,\sigma}
                               \rme^{-\rmi k\cdot x}$.
Then, noting two commutation relations
\begin{equation}
 \left[\hat{a}_{-k,\dn} \hat{a}_{k,\up},
                \left(\sum_{p\in \calK} 
                  \left(\hat{d}_{p,\up}^\dagger\hat{a}_{-p,\dn}^\dagger
                    + \hat{a}_{p,\up}^\dagger \hat{d}_{-p,\dn}^\dagger\right)
     \right)\right]
 = -\left(\hat{d}_{k,\up}^\dagger\hat{a}_{k,\up}+
             \hat{d}_{-k,\dn}^\dagger\hat{a}_{-k,\dn}
                      \right)
\end{equation}
and
\begin{equation}
 \left[-\left(\hat{d}_{k,\up}^\dagger\hat{a}_{k,\up}+
             \hat{d}_{-k,\dn}^\dagger\hat{a}_{-k,\dn}
                      \right) ,
                \left(\sum_{p\in \calK} 
                  \left(\hat{d}_{p,\up}^\dagger\hat{a}_{-p,\dn}^\dagger
                    + \hat{a}_{p,\up}^\dagger \hat{d}_{-p,\dn}^\dagger\right)
     \right)\right]
 = -2\hat{d}_{k,\up}^\dagger \hat{d}_{-k,\dn}^\dagger, 
\end{equation}  
we obtain
\begin{eqnarray}
 \PhiG 
    &=&
   \frac{1}{N!}
              \calP_D\left(\sum_{k\in\calK}(\cos k_2-\cos k_1)
                \hat{d}_{k,\up}^\dagger \hat{d}_{-k,\dn}^\dagger
              \right)^\Np
              \left(\sum_{x\in D} \psi_{x}^\dagger 
               \right)^N\Phi_0 \nonumber\\ 
  &=& \frac{\Np!}{N!}
              \calP_D
               \left(
                  \sum_{B\in C(\calB) \calB;|B|=\Np}
                      \prod_{\{x,y\}\in B}(\phi_{\{x,y\}}^d)^\dagger
                \right)
              \left(\sum_{x\in D} \psi_{x}^\dagger 
               \right)^N\Phi_0 
\label{eq:PhiG-hole-representation}
\end{eqnarray}
with
\begin{equation}
 (\phi_{\{x,y\}}^d)^\dagger = \frac{1}{2}\rme^{-\rmi\pi \delta^1\cdot(x-y)}
                    (d_{x,\up}^\dagger d_{y,\dn}^\dagger+d_{y,\up}^\dagger d_{x,\dn}^\dagger).
\end{equation}
To get the second line in \eqref{eq:PhiG-hole-representation}
we used $\calP_D d_{x,\up}^\dagger d_{x,\dn}^\dagger=0$.
The operator $(\phi_{\{x,y\}}^d)^\dagger$
corresponds to a two-hole singlet state formed by holes at the Cu-sites.
From expression \eqref{eq:PhiG-hole-representation} 
we find that the state $\PhiG$ can be regarded also as a projected state
in which the Zhang-Rice singlets condense and the remaining holes at
the Cu-sites are forming nearest-neighbour singlet states.

Here it should be noted that, despite of the
form~\eqref{eq:PhiG-hole-representation}, $\PhiG$ does not exhibit 
long-range order associated with the Zhang-Rice singlets.
In fact, it is easy to see that
\begin{equation}
 \expec{\PhiG,\psi_{x}^\dagger \psi_{y}\PhiG}=0
\end{equation}
for $x\ne y$, since there is no charge fluctuation on the Cu-sites.
It is important to consider  long-range order associated with 
movable holes on the O-sites, which we discuss in the next section.  
%%%%%%%%%%%%%%%%%%%%%%%%%%%%%%%%%%%%%%%
\section{Order Parameter}
\label{sc:Order Parameter}
In this section we estimate
the value of a $d$-wave order parameter for the state
\eqref{eq:pairing-state}. Let
\begin{equation}
\Delta=\frac{1}{|D|}
\sum_{\{x,y\}\in \calB}\phi_{\{x,y\}}^a
=\frac{1}{|D|}\zeta.
\label{eq:Delta}
\end{equation}
We then define 
\begin{eqnarray}
\label{eq:mu}
 \mu_{\La,N}&=&\sqrt{\frac{\expPhiG{\Delta^\dagger\Delta}}
                      {\expPhiG{}}}, \\
 \mu_\delta&=&\limtwo{|D|,N\to\infty}{N/|D|=\delta}\mu_{\La,N}, 
\end{eqnarray}
where the limit is taken with $N/|D|$ kept fixed to $\delta$.
This order parameter measures a long range correlation between 
spin-singlet pairs corresponding $\phi_{\{x,y\}}^a$.
 
We firstly show that 
\begin{equation}
\expPhiG{\Delta^\dagger \Delta}
        =\frac{\Np+1}{2|D|^2}\sum_{x\in D}\sum_{\sigma=\up,\dn}
                {\expPhiG{a_{x,\sigma}^\dagger 
                          a_{x,\sigma}b_{x,-\sigma}^\dagger b_{x,-\sigma}}}
\label{eq:order-parameter}
\end{equation}
which is crucial for our estimation of $\mu_{\delta}$ (recall $\Np=(|D|-N)/2$).
To see this, we observe that 
\begin{eqnarray}
 \expec{ (\zeta)^\Np \Psi_0,a_{x,\up}^\dagger b_{x,\dn}^\dagger(\zeta)^{\Np+1}\Psi_0}  
&=& \expec{ (\zeta)^\Np d_{x,\dn}^\dagger \prod_{y\in D\backslash\{x\}} \psi_{y}^\dagger \Phi_0, 
            b_{x,\dn}^\dagger(\zeta)^{\Np+1}\Psi_0}\nonumber\\
&=& (\Np+1)\expec{ (\zeta)^\Np d_{x,\dn}^\dagger \prod_{y\in D\backslash\{x\}} \psi_{y}^\dagger \Phi_0, 
            b_{x,\dn}^\dagger b_{x,\dn} (\zeta)^\Np d_{x,\dn}^\dagger 
                          \prod_{y\in D\backslash\{x\}} \psi_{y}^\dagger\Phi_0}\nonumber\\
&=& (\Np+1)\expec{ (\zeta)^\Np \Psi_0, 
            a_{x,\up}^\dagger a_{x,\up} b_{x,\dn}^\dagger b_{x,\dn} (\zeta)^\Np\Psi_0}.
\label{eq:expec1}
\end{eqnarray}
To get the second line we used the commutation relation 
\begin{equation} 
    [\zeta a_{x,\sigma}^\dagger,~a_{x,\sigma}^\dagger \zeta]=\sgn[\sigma]b_{x,-\sigma},
\label{eq:commutation-a-zeta}
\end{equation}
which immediately follows from the real space representation \eqref{eq:zeta3} of $\zeta$.
Then, by using the spin-rotation symmetry for $\PhiG$,  
\eqref{eq:order-parameter} follows from \eqref{eq:Delta} and \eqref{eq:expec1}.

By noting the inequality  
\begin{eqnarray}
\expPhiG{a_{x,\sigma}^\dagger a_{x,\sigma} b_{x,-\sigma}^\dagger b_{x,-\sigma}}
& = & \expPhiG{a_{x,\sigma}^\dagger a_{x,\sigma}(1-b_{x,-\sigma}b_{x,-\sigma}^\dagger)}  \nonumber\\
& \le &  \expPhiG{a_{x,\sigma}^\dagger a_{x,\sigma}}
\label{eq:bound-aabb} 
\end{eqnarray}
and the fact that the number of $a$-holes (which are holes in the state
corresponding to the $a$-operators) is exactly $N$ for $\PhiG$,
we find that
$\mu_{\La,N}$ is bounded from above as 
\begin{equation}
 \mu_{\La,N} \le \sqrt{\left(\frac{\Np+1}{|D|}\right)\left(\frac{N}{2|D|}\right)}
\end{equation}
The limit  is thus bounded from above as 
\begin{equation}
 \mu_{\delta}\le\frac{1}{2}\sqrt{\delta(1-\delta)}.
\end{equation}

As for a lower bound for $\mu_\delta$ we have the following result.
Let $\frac{8}{9}\le\delta\le1$. Then we have that 
\begin{equation}
 \mu_\delta \ge \frac{1}{2}\sqrt{\gamma_\delta I(\delta)(1-\delta)},
\label{eq:mu-lower-bound}
\end{equation}
where $\gamma_\delta=\frac{9\delta-8}{2(8\delta-7)}$ and 
\begin{equation}
 I(\delta)= \frac{2}{(2\pi)^2}\int_{|k_i|\le\pi} \epsilon_b(k) 
                         \chi[\epsilon_b(k)\le \epsilon_\delta] \mathrm{d}k
\end{equation}
with $\epsilon_b(k)=(\cos k_1-\cos k_2)^2$.
Here $\epsilon_\delta$ is determined by
\begin{equation}
  \delta=\frac{2}{(2\pi)^2}\int_{|k_i|\le\pi}  
                         \chi[\epsilon_b(k)\le \epsilon_\delta] \mathrm{d}k.
\end{equation}
The inequality \eqref{eq:mu-lower-bound} means that the state $\PhiG$
exhibits ODLRO with $d$-wave symmetry for $\frac{8}{9}< \delta < 1$.
The calculation for this bound is somewhat complicated and technical.
We defer the proof to section \ref{s:Estimation-of-a-lower-bound}.
It should be noted that the above lower bound is not optimal at all and
never means that there is no $d$-wave pairing order in a low density
region of holes.
It is desirable to obtain an improved bound in the future.
%%%%%%%%%%%%%%%%%%%%%%%%%%%%%%%%%%%%%%%%
\section{Hamiltonian with Ground State $\PhiG$}
\label{sc:Hamiltonian}
So far we have constructed the pairing state $\PhiG$ with $d$-wave symmetry and
studied its properties.
In this section we propose a Hamiltonian $H$ on $\La$ whose ground state is given
by~$\PhiG$.

Let us define the number operator $n_{x,\sigma}^a$, with
$\sigma=\up,\dn$,
and the spin operators $S_{x,\alpha}^{a}$, with $\alpha=1,2,3$, 
corresponding to the $a$-operators by
\begin{eqnarray}
 n_{x,\sigma}^a&=&a_{x,\sigma}^\dagger a_{x,\sigma},\\
 S_{x,1}^a     &=& \frac{1}{2}(a_{x,\up}^\dagger a_{x,\dn}+a_{x,\dn}^\dagger a_{x,\up}),\\ 
 S_{x,2}^a     &=& \frac{1}{2\rmi}(a_{x,\up}^\dagger a_{x,\dn}-a_{x,\dn}^\dagger a_{x,\up}),\\ 
 S_{x,3}^a     &=& \frac{1}{2}(a_{x,\up}^\dagger a_{x,\up}-a_{x,\dn}^\dagger a_{x,\dn}).
\end{eqnarray}
We also define 
\begin{equation}
 n_x^a=n_{x,\up}^a+n_{x,\dn}^a.
\end{equation}
The number and the spin operators for the $b$- and the $d$-operators are defined
similarly.
By using these operators, the Hamiltonian $H$ is defined as follows:
\begin{equation}
  H=H_0+H_1
 \label{eq:Hamiltonian}
\end{equation} 
with 
\begin{eqnarray}
\label{eq:H0}
 H_0 & = & -\ved\sum_{x\in D} n_{x}^d + U\sum_{x\in} n_{x,\up}^d n_{x,\dn}^d
               +J_0 \sum_{x\in D}\mbox{\boldmath $S$}_{x}^a\cdot \mbox{\boldmath $S$}_{x}^d , \\
 H_1 & = & \frac{3}{4}J_1 \sum_{x\in D}\sum_{\sigma=\up,\dn} 
                     \left(a_{x,\sigma}^\dagger a_{x,\sigma}
                           +b_{x,\sigma}^\dagger b_{x,\sigma}\right)
      \nonumber \\
     &&     +J_1 \sum_{x\in D}\left(
                  \mbox{\boldmath $S$}_{x}^a\cdot \mbox{\boldmath $S$}_{x}^d
                 +\mbox{\boldmath $S$}_{x}^b\cdot \mbox{\boldmath $S$}_{x}^d  
                 +\mbox{\boldmath $S$}_{x}^a\cdot \mbox{\boldmath $S$}_{x}^b 
                         -\frac{3}{4}n_x^a\cdot n_x^b
                   \right).
\label{eq:H1}
\end{eqnarray}
Here, all the parameters, $\ved,~U,~J_0$ and $J_1$, are positive,
and $\varepsilon_d$ is assumed 
to take values in  $\frac{3}{4}J_0<\ved<\frac{3}{4}J_0+U$.
It should be noted that one can rewrite $H$ by using the $d$- and
the $p$-operators, although it has a somewhat complicated form.
It is also noted that we do not take any  peculiar limit, 
such as $U\to\infty$ and $J_0\to\infty$, 
and thus $H$ acts on a whole Hilbert space constructed 
by the $d$- and the $p$-operators.

We shall show that the lowest energy of $H_0$ for the hole number
$\Nh=|D|+N$ with $0< N \le |D|$ is
$\vezero=-\ved|D|-\frac{3}{4}J_0 N$, which is attained by 
the Zhang-Rice singlet states in $\HZRS$.

Let $\Nhd$ be the eigenvalue of $\sum_{x\in D}n_{x}^d$, the number
of $d$-holes.
Since $\Nhd$ is a conserved quantity for $H_0$, it is convenient 
to  decompose the $\Nh$-hole Hilbert space 
into the subspaces with fixed $\Nhd$.
We denoted by
$\mathbf{H}_\Nhd^\Nh$ the subspace with fixed $\Nhd$ and by 
$E(\Nhd)$ the lowest energy of $H_0$ for the states in $\mathbf{H}_\Nhd^\Nh$.

Let us examine each term in $H_0$.
The eigenvalue of the first sum in $H_0$ is $-\ved\Nhd$ for the
states in $\mathbf{H}_\Nhd^\Nh$.
The lowest eigenvalue for the second sum is zero which is attained  by 
the states without doubly occupied $d$-states.    
The eigenvalues of $J_0\mbox{\boldmath $S$}_{x}^a\cdot \mbox{\boldmath
$S$}_{x}^d$
are $-\frac{3}{4}J_0$, $0$ and $\frac{1}{4}J_0$. 
We have eigenvalue $-\frac{3}{4}J_0$ 
when each of the $d$-state and the $a$-state at site $x$ is occupied by one hole 
and furthermore the two holes in these states form the spin-singlet state.    

It immediately follows from the above observation that  
$E(|D|)=\vezero$, which is attained by the states in 
$\HZRS\subset\mathbf{H}_{|D|}^\Nh$. 
In the case $0\le\Nhd<|D|$, 
noting that there are $\Nhp=|D|+N-\Nhd$ holes on the O-sites, we have
\begin{eqnarray}
E(\Nhd)&=&-\ved\Nhd-\frac{3}{4}J_0\min(\Nhd,\Nhp)
\nonumber\\
       &>& \vezero + \frac{3}{4}J_0\left\{\Nhp-\min(\Nhd,\Nhp)\right\}
\nonumber\\
       &\ge& \vezero.  
\end{eqnarray}
Here the second line follows from the assumptions $0<\frac{3}{4}J_0<\ved$
and $\Nhd<|D|$(or $N<\Nhp$), and the third line follows from $\Nhp\ge\min(\Nhd,\Nhp)$.
In the case $|D|<\Nhd\le|D|+N$, noting that there are, at least,
$(\Nhd-|D|)$ doubly occupied $d$-states, we have
\begin{eqnarray}
E(\Nhd)&=&-\ved\Nhd+U(\Nhd-|D|)-\frac{3}{4}J_0 \Nhp
\nonumber\\
       &=& \vezero + \left(\frac{3}{4}J_0+U-\ved \right)(\Nhd-|D|)
\nonumber\\
       &>& \vezero.  
\end{eqnarray}
Here the final inequality follows from the assumptions 
$\ved<\frac{3}{4}J_0+U$ and $|D|<\Nhd$.
As a result, we have $E(\Nhd)>\vezero$ for $\Nhd\ne |D|$, 
which proves the claim.   

We have shown that the lowest-energy states of $H_0$ are 
the Zhang-Rice singlet states in $\HZRS$.
In the following we shall show that $H_1$ is positive semi-definite and
$\PhiG$ in $\HZRS$ is its zero energy state.
This implies $H=H_0+H_1\ge \vezero$ and $H\PhiG=\vezero\PhiG$.
We thus conclude that $\PhiG$ is a ground state of $H$.

By a straightforward but somewhat lengthy calculation, one finds that 
$H_1$ is rewritten as follows:
\begin{equation}
 H_1=\frac{3}{8}J_1\sum_{x\in D}\sum_{m=1}^2\sum_{l=1}^4 
          \left[(K_{x,l}^m)^\dagger K_{x,l}^m + K_{x,l}^m(K_{x,l}^m)^\dagger \right]
\label{eq:H11}
\end{equation} 
with
\begin{eqnarray}
   K_{x,1}^1 & = &  b_{x,\up}^\dagger a_{x,\dn} d_{x,\dn} , \\
   K_{x,2}^1 & = &  \frac{1}{\sqrt{3}}
                 \left(
                    b_{x,\up}^\dagger a_{x,\dn} d_{x,\up} 
                   +b_{x,\up}^\dagger a_{x,\up} d_{x,\dn} 
                   -b_{x,\dn}^\dagger a_{x,\dn} d_{x,\dn}  
                  \right),\\
   K_{x,3}^1 & = &  \frac{1}{\sqrt{3}}
                 \left(
                    b_{x,\up}^\dagger a_{x,\up} d_{x,\up} 
                   -b_{x,\dn}^\dagger a_{x,\dn} d_{x,\up} 
                   -b_{x,\dn}^\dagger a_{x,\up} d_{x,\dn}  
                  \right),\\
   K_{x,4}^1 & = &  -b_{x,\dn}^\dagger a_{x,\up} d_{x,\up}, 
\end{eqnarray} 
and 
\begin{eqnarray}
   K_{x,1}^2 & = &  a_{x,\up}^\dagger b_{x,\dn} d_{x,\dn} , \\
   K_{x,2}^2 & = &  \frac{1}{\sqrt{3}}
                 \left(
                    a_{x,\up}^\dagger b_{x,\dn} d_{x,\up} 
                   +a_{x,\up}^\dagger b_{x,\up} d_{x,\dn} 
                   -a_{x,\dn}^\dagger b_{x,\dn} d_{x,\dn}  
                  \right),\\
   K_{x,3}^2 & = &  \frac{1}{\sqrt{3}}
                 \left(
                    a_{x,\up}^\dagger b_{x,\up} d_{x,\up} 
                   -a_{x,\dn}^\dagger b_{x,\dn} d_{x,\up} 
                   -a_{x,\dn}^\dagger b_{x,\up} d_{x,\dn}  
                  \right),\\
   K_{x,4}^2 & = &  -a_{x,\dn}^\dagger b_{x,\up} d_{x,\up}. 
\end{eqnarray} 
It follows from this representation that $H_1$ is positive
semi-definite.
Therefore, the lowest energy of $H_1$ is greater than or equal to zero, and 
any zero energy state $\Phi$ of $H_1$, if it exists, must satisfy
$(K_{x,l}^m)^\dagger\Phi=0$ and $K_{x,l}^m\Phi=0$ for all $m=1,2$,
$l=1,\dots,4$ and $x\in D$.
We shall prove that $\PhiG$ indeed satisfies these conditions.

We start with the case of $m=1$ and $l=1$.
By using the commutation relation~\eqref{eq:commutation-a-zeta}, we
have
\begin{equation}
 (K_{x,1}^1)^\dagger\zeta
        =d_{x,\dn}^\dagger a_{x,\dn}^\dagger b_{x,\up}\zeta
             =d_{x,\dn}^\dagger (b_{x,\up})^2
              +\zeta d_{x,\dn}^\dagger a_{x,\dn}^\dagger b_{x,\up}
             =\zeta (K_{x,1}^1)^\dagger.
\end{equation}
This together with $(K_{x,1}^1)^\dagger \psi_x^\dagger=0$, which follows from
$(a_{x,\sigma}^\dagger)^2=(d_{x,\sigma}^\dagger)^2=0$, 
leads to 
\begin{equation}
 (K_{x,1}^1)^\dagger\left(\zeta\right)^\Np\Psi_0
=\left(\zeta\right)^\Np (K_{x,1}^1)^\dagger\Psi_0=0.
\end{equation}
{From}
$a_{x,\dn}d_{x,\dn}\psi_{x}^\dagger
  =-d_{x,\dn}a_{x,\up}^\dagger d_{x,\dn}^\dagger a_{x,\dn} 
     + a_{x,\dn}d_{x,\up}^\dagger a_{x,\dn}^\dagger d_{x,\dn}$
we immediately obtain 
\begin{equation}
 K_{x,1}^1 \left(\zeta\right)^\Np\Psi_0
= b_{x,\up}^\dagger \left(\zeta\right)^\Np a_{x,\dn}d_{x,\dn}\Psi_0=0.
\end{equation}
We thus conclude $(K_{x,1}^1)^\dagger\PhiG=K_{x,1}^1\PhiG=0$ for all $x$ in $D$.

Let us consider the cases of $m=1$ and $l=2,3,4$.
Define spin-lowering and raising operators as
\begin{eqnarray}
  S^- &=& \sum_{x\in D}(a_{x,\dn}^\dagger a_{x,\up}+d_{x,\dn}^\dagger d_{x,\up}),\\ 
  S^+ &=& \sum_{x\in D}(a_{x,\up}^\dagger a_{x,\dn}+d_{x,\up}^\dagger d_{x,\dn}). 
\end{eqnarray}
{From} the results for $l=1$ we have $S^+(K_{x,1}^1)^\dagger\PhiG=0$.
It is easy to see that 
$S^+(K_{x,1}^1)^\dagger=\sqrt{3}(K_{x,2}^1)^\dagger+(K_{x,1}^1)^\dagger S^+$, 
$S^+(K_{x,2}^1)^\dagger=      2 (K_{x,3}^1)^\dagger+(K_{x,2}^1)^\dagger S^+$, 
and
$S^+(K_{x,3}^1)^\dagger=\sqrt{3}(K_{x,4}^1)^\dagger+(K_{x,3}^1)^\dagger S^+$. 
Substituting the first relation into $S^+(K_{x,1}^1)^\dagger\PhiG=0$ and
noting $S^+\PhiG=0$, we find $(K_{x,2}^1)^\dagger\PhiG=0$.
Repeating the same argument, we have $(K_{x,3}^1)^\dagger\PhiG=(K_{x,4}^1)^\dagger\PhiG=0$.
By using 
$S^-K_{x,1}^1\PhiG=0$, 
$S^-K_{x,1}^1=-\sqrt{3}K_{x,2}^1+K_{x,1}^1 S^-$, 
$S^-K_{x,2}^1=-2K_{x,3}^1+K_{x,2}^1 S^-$, 
$S^-K_{x,3}^1=-\sqrt{3}K_{x,4}^1+K_{x,3}^1 S^-$,
and $S^-\PhiG=0$,  
we similarly obtain $K_{x,l}^1\PhiG=0$ for $l=2,3,4$.

Proceeding in the same way,  we obtain 
$(K_{x,l}^2)^\dagger\Phi=K_{x,l}^2\Phi=0$ for $l=1,\dots,4$ and $x\in D$.
This completes the proof of the claim.

We remark that the uniqueness of the ground state of $H$ for
each hole number is not proved at present.
We hope that this will be clarified in a future study.
%%%%%%%%%%%%%%%%%%%%%%%%%%%%%%%%%%%%%%%%
\section{Estimation of a Lower Bound for $\mu_\delta$}
\label{sc:Lower Bound}
\label{s:Estimation-of-a-lower-bound}
In this section we estimate a lower bound for $\mu_\delta$.
We will show later that 
\begin{equation}
 \frac{\expPhiG{a_{x,\sigma}^\dagger a_{x,\sigma} b_{x,-\sigma}^\dagger b_{x,-\sigma}}}
      {\expPhiG{b_{x,-\sigma}^\dagger b_{x,-\sigma}}} \ge \gamma_{\La,N},
\label{eq:bound1}
\end{equation}
with $\gamma_{\La,N}=\frac{9N-8|D|}{2(8N-7|D|)}$, for $N\ge(8|D|)/9$.
It follows from this inequality that
\begin{equation}
 \mu_{\La,N}\ge \sqrt{\gamma_{\La,N}\frac{\Np+1}{2|D|^2} 
 \frac{\sum_{x,\sigma}\expPhiG{ b_{x,\sigma}^\dagger b_{x,\sigma}}}
      {\expPhiG{}}}.
\label{eq:mu-finite-bound}
\end{equation}
Here, we have that
\begin{equation}
 \sum_{x\in D}\sum_{\sigma=\up,\dn}
   \expPhiG{ b_{x,\sigma}^\dagger b_{x,\sigma}}
=
 \sum_{k\in\calK}\sum_{\sigma=\up,\dn}
         \expPhiG{ \epsilon_b(k)\hat{a}_{k,\sigma}^\dagger \hat{a}_{k,\sigma}},
\end{equation} 
and it is easy to find that the right-hand-side is bounded from below by
\begin{equation}
 {2}\sum_{l=1}^{N/2} \epsilon_b(k^{(l)})\expPhiG{},
\end{equation}
where $\epsilon_b(k^{(1)}) \le \epsilon_b(k^{(2)}) \dots \le \epsilon_b(k^{(|D|)})$
is an arrangement of $\epsilon_b(k)$ with $k\in\calK$ in ascending order.
Substituting this lower bound into \eqref{eq:mu-finite-bound} and taking the limit, we obtain 
\eqref{eq:mu-lower-bound}.

In what follows we prove  inequality \eqref{eq:bound1}.
By the spin-rotation symmetry for $\PhiG$, it suffices to consider the
case of $\sigma=\up$.
By the translation symmetry,
we can also assume $x\in \De$ without loss of generality .
We first show that the left-hand-side of \eqref{eq:bound1} with $\sigma=\up$ is rewritten as
\begin{equation}
\frac{\sum_{S\subset\De;x\notin S, |S|=\Np}W_x(S)}
     {2\sum_{S\subset\De;|S|=\Np}W_x(S)}
=
\frac{\sum_{S\subset\De;x\notin S, |S|=\Np}W_x(S)}
     {2(\sum_{S\subset\De;x\notin S,|S|=\Np}W_x(S)+\sum_{S\subset\De;x\in S,|S|=\Np}W_x(S))}
\label{eq:weight-representation}
\end{equation}
with the nonnegative weights
\begin{equation}
  W_x(S)=\left<
   \left(\prod_{z\in S}\tphiz\right)\Psi_0,
              b_{x,\dn}^\dagger b_{x,\dn}
   \left(\prod_{z\in S}\tphiz\right)\Psi_0
   \right>,
\end{equation}
where $\tphiz=(b_{z,\dn}a_{z,\up}+a_{z,\dn}b_{z,\up})$.
To see this, note that
\begin{equation}
 \zeta=\sum_{z\in \De} b_{z,\dn}a_{z,\up}+\sum_{z \in \Do}b_{z,\dn}a_{z,\up}
      =\sum_{z\in \De} (b_{z,\dn}a_{z,\up}+a_{z,\dn}b_{z,\up})
      =\sum_{z\in \De} \tphiz.
\end{equation}
Then, since $(\tphiz)^2\Psi_0=0$ (which follows from
$a_{z,\dn}a_{z,\up}\Psi_0=0$), 
$\PhiG$ is expanded as
\begin{equation}
 \PhiG=\Np!\sumtwo{S\subset \De}{|S|=\Np}\left(\prod_{z\in S}\tphiz\right)\Psi_0.
\label{eq:PhiG-expansion}
\end{equation}
Since $a_{x,\up}\left(\prod_{z\in S}\tphiz\right)\Psi_0=0$ for $x\in S$ (which again
follows from $a_{z,\dn}a_{z,\up}\Psi_0=0$), and 
\begin{equation}
 \left<\left(\prod_{z\in S^\prime}\tphiz\right)\Psi_0, 
          a_{x,\up}^\dagger a_{x,\up} b_{x,\dn}^\dagger b_{x,\dn}
          \left(\prod_{z\in S}\tphiz\right)\Psi_0 
  \right>=0~~\mbox{for $S^\prime\ne S$},
\end{equation}
we have that
\begin{eqnarray} 
\expPhiG{a_{x,\up}^\dagger a_{x,\up} b_{x,\dn}^\dagger b_{x,\dn}}
&&=(\Np!)^2\hspace*{-7pt}\sumtwo{S\subset \De}{x\notin S, |S|=\Np}\hspace{-7pt} 
   \left<
   \left(\prod_{z\in S}\tphiz\right)\Psi_0,
       ~a_{x,\up}^\dagger a_{x,\up} b_{x,\dn}^\dagger b_{x,\dn}
   \left(\prod_{z\in S}\tphiz\right)\Psi_0
   \right>. \nonumber\\
&& = \frac{1}{2}(\Np!)^2\hspace*{-7pt}\sumtwo{S\subset \De}{x\notin S, |S|=\Np} W_x(S).
\label{eq:weight-representation1}
\end{eqnarray}
To get the second line we used
\begin{equation} 
 \left<\psi_{x}^\dagger\Phi_0,~a_{x,\up}^\dagger a_{x,\up}\psi_x^\dagger\Phi_0\right>
  =1=\left<\psi_{x}^\dagger\Phi_0,~\psi_x^\dagger\Phi_0\right>/2.
 \label{eq:half}
\end{equation}
Likewise, we have that 
\begin{equation}
 \expPhiG{b_{x,\dn}^\dagger b_{x,\dn}}=(\Np!)^2\sum_{S\subset \De;|S|=\Np} W_x(S),
\end{equation}
which together with \eqref{eq:weight-representation1} leads to \eqref{eq:weight-representation}.

Before proceeding, we need to introduce some notation. 
For each $z\in \De$, define 
$D_{\mathrm{o},z}=\{y~|~|y-z|=1,~y\in\Do\}$, which is the collection of
the  nearest neighbour sites of $z$.
We say that $z$ and $z^\prime$ in $\De$ are connected if 
$D_{\mathrm{o},z}\cap D_{\mathrm{o},z^\prime}\ne\emptyset$.
For $S\subset\De$ which does not contain $x$,
we call $z$ an isolated point in $S$ if $z$ is not connected any other
sites in $S\cup\{x\}$, and write $D_x(S)$ for the collection of these isolated
points in $S$.
It is noted that, if $y\in D_{x}(S^\prime\cup\{y\})$,
the weight $W_x(S^\prime\cup\{y\})$ is reduced as 
\begin{equation}
 W_x(S^\prime\cup\{y\})=\frac{1}{2}W_x(S^\prime),
\label{eq:weight}
\end{equation}
since $a_{y^\prime,\sigma}^\dagger$ with $|y^\prime-y|\le1$ commutes with $b_{x,\dn}^\dagger
b_{x,\dn}\prod_{z\in S^\prime}\tphiz$ and thus
\begin{equation}
 \left<
  \Psi_0,\tilde{\phi}_y^\dagger \tilde{\phi}_y \Psi_0
 \right>
=\frac{1}{4}\sumtwo{y^\prime\in\Do}{|y^\prime-y|=1}\sum_{\sigma=\up,\dn}
 \left<
  \Psi_0,a_{y,\sigma}^\dagger a_{y^\prime,-\sigma}^\dagger  
         a_{y^\prime,-\sigma}         a_{y,\sigma}          \Psi_0
 \right>
=\frac{1}{2} \left<
  \Psi_0, \Psi_0
 \right>.
\end{equation}
(Recall \eqref{eq:half}.) 
We denote by $\calD_x(\Np,l)$ the collection of subsets $S$ of $\De$ such that
$x\notin S$, $|S|=\Np$ and $|D_x(S)|=l$. 

Since the value of $|D_x(S)|$ is determined for each $S\subset
\De$, we have  
\begin{equation}
 \sumtwo{S\subset\De}{x\notin S, |S|=\Np}W_x(S)
=\sum_{l=0}^{\Np}\sum_{S\in\calD_x(\Np,l)}W_x(S)
\ge
\sum_{l=1}^{\Np}\sum_{S\in\calD_x(\Np,l)}W_x(S).
\label{eq:bound2} 
\end{equation}
Now fix $l\ge1$. Noting that there are $l$ isolated sites in $S\in\calD_x(\Np,l)$,
we find
\begin{eqnarray}
\sum_{S\in\calD_x(\Np,l)}W_x(S) 
   &=& \frac{1}{l}\sum_{S\in\calD_x(\Np,l)}
        \sum_{y\in \De}W_x(S)\chi[y\in D_x(S)] \nonumber\\
   &=& 
     \frac{1}{l}\sum_{S\in\calD_x(\Np,l)}
         \sum_{y\in \De}\sum_{S^\prime\in\calD_x(\Np-1,l-1)}W_x(S)\chi[y\in D_x(S)]
                                                     \chi[S^\prime=S\backslash\{y\}] \nonumber\\
   &=&\frac{1}{2l}\sum_{S^\prime\in\calD_x(\Np-1,l-1)}\hspace*{-10pt}W_x(S^\prime)\hspace*{-5pt}
                  \sum_{S\in\calD_x(\Np,l)}\sum_{y\in \De}\chi[y\in D_x(S)]
                                                     \chi[S=S^\prime\cup\{y\}] \nonumber\\
   &\ge& \frac{1}{2\Np}\left(\frac{|D|}{2}-9\Np\right)
              \sum_{S^\prime\in\calD_x(\Np-1,l-1)}\hspace*{-10pt}W_x(S^\prime)
\label{eq:bound3}
\end{eqnarray}
To get the second line, note that  removing an isolated point in $S\in\calD_x(\Np,l)$
yields an element in $\calD_x(\Np-1,l-1)$.
The third line follows from \eqref{eq:weight}.
The last inequality is obtained as follows. Each site $z$ in $\De$ has
8 connected sites.
Therefore, for every $S^\prime\in\calD_x(\Np-1,l-1)$, 
there exist at least $|\De|-9\Np$ sites, $y$, such that
$y$ is an isolated point in $S^\prime\cup\{y\}$, and  
$S^\prime\cup\{y\}$ becomes an element in $\calD_x(\Np,l)$. 
Note that $|\De|-9\Np$ is a positive number  by the assumption.
Then, by using $l\le\Np$, we get the last inequality.

From \eqref{eq:bound2} and \eqref{eq:bound3} we get
\begin{equation}
 \sumtwo{S\subset\De}{x\notin S, |S|=\Np}W_x(S)
 \ge \frac{1}{2\Np}\left(\frac{|D|}{2}-9\Np\right)
  \sum_{l=0}^{\Np-1}
              \sum_{S\in\calD_x(\Np-1,l)}\hspace*{-10pt}W_x(S).
\label{eq:bound4}
\end{equation}
Here, for $x\notin S$, we have
\begin{eqnarray}
 W_x(S\cup\{x\}) &=& \left<\left(\prod_{z\in S}\tphiz\right)\Psi_0, 
            a_{x,\dn}^\dagger a_{x,\dn}b_{x,\up}^\dagger b_{x,\up}b_{x,\dn}^\dagger b_{x,\dn}
                       \left(\prod_{z\in S}\tphiz\right)\Psi_0\right> \nonumber\\
               &=&\frac{1}{2}\left<\left(\prod_{z\in S}\tphiz\right)\Psi_0,  
                        (1 -b_{x,\up} b_{x,\up}^\dagger)b_{x,\dn}^\dagger b_{x,\dn}
                       \left(\prod_{z\in S}\tphiz\right)\Psi_0\right> \nonumber\\
               &\le& \frac{1}{2}W_x(S).
\end{eqnarray}
It follows from this inequality and \eqref{eq:bound4} that
\begin{eqnarray}
 \sumtwo{S\subset\De}{x\notin S, |S|=\Np}W_x(S)
 &\ge& \frac{1}{\Np}\left(\frac{|D|}{2}-9\Np\right)
  \sum_{l=0}^{\Np-1}
              \sum_{S\in\calD_x(\Np-1,l)}\hspace*{-10pt}W_x(S\cup\{x\}) \nonumber\\
 &=&\left(\frac{|D|}{2\Np}-9\right)\sumtwo{S\subset\De}{x\in S,|S|=\Np}W_x(S).
\label{eq:bound5}
\end{eqnarray}
From \eqref{eq:weight-representation} and \eqref{eq:bound5} we obtain
the desired inequality~\eqref{eq:bound1}.
%%%%%%%%%%%%%%%%%%%%%%%%%%%%%%%%%%%%%%%%%%%%%%%%%%%%%%%%%%%%%%%%%%%%%%%%%%%%%%%%%%%
\section{Summary and Remarks}
In this paper, for the even numbers $\Nh$ of holes in $|D|<\Nh\le2|D|$,
we have constructed a pairing state $\PhiG$ with $d$-wave symmetry
which is expanded in terms of the Zhang-Rice singlet states.
We have calculated upper and lower bounds of the ODLRO parameter for
$\PhiG$ as a function of the hole concentration.
We have also presented the concrete Hamiltonian $H=H_0+H_1$~\eqref{eq:Hamiltonian}
on the CuO$_2$ plain which has $\PhiG$ as its ground state.
We have proved that the lowest energy states of $H_0$~\eqref{eq:H0} are the Zhang-Rice
singlet states and then have shown that, by using the
positive-semidefiniteness of $H_1$~\eqref{eq:H11},
the pairing state $\PhiG$ consisting
of the Zhang-Rice singlet states attains the ground state energy of the
whole Hamiltonian $H$.
The uniqueness of the ground state is not proved at present, and
we leave this as a problem in a future study.

It is noted that $H_0$ with $J_0=0$ becomes the Hamiltonian of the $d$-$p$
(or 3-band) model in the atomic
limit~\cite{Emery87,Hirsch87,ZhangRice88}, 
and $H_0$ with $J_0\ne0$ is
essentially the same as the effective Hamiltonian derived by taking into
account the hopping terms between Cu- and O-sites as a perturbation in
the limit~\cite{ZhangRice88}.
The idea of the Zhang-Rice singlet is based on
this effective Hamiltonian, and the \tJmodel~is obtained by 
furthermore considering the motion of the Zhang-Rice singlets
perturbatively with the inclusion of the antiferromagnetic interactions
between Cu-holes 
\begin{equation}
 H_2=J_2\sum_{x,y\in D;|x-y|=1}\vecS_{x}^d\cdot\vecS_{y}^d,
\end{equation}
which is the effective interaction due to the hopping process between
neighbouring Cu-sites~\cite{ZhangRice88}.

In the $|D|$-hole case, the present Hamiltonian has degenerate
paramagnetic ground states with one hole per Cu-site 
and does not exhibit antiferromagnetism
which is essential to high-$\Tc$ cuprates.
This will be improved if we consider the modified Hamiltonian 
$H_0+H_1+H_2$. 
This Hamiltonian or more generally the $d$-$p$ Hamiltonian with $H_1$
may be able to reproduce the essential features of high-$\Tc$ cuprates,
such as antiferromagnetism at low doping concentrations 
and charge density order (or a stripe structure) 
between the antiferromagnetic and the
superconducting states.
We believe that 
further investigations about modified models based on our Hamiltonian
which is now shown to exhibit ODRLO with $d$-wave symmetry 
will contribute the understanding of high-$\Tc$ cuprate superconductivity.  
%%%%%%%%%%%%%%%%%%%%%%%%%%%%%%%%%%%%%%%%%%%%%%%%%%%%%%%%%%%%%%%%%%%%%%%%%%%%%%%%%%%
\bigskip\\
\textbf{\Large Acknowledgements}
\bigskip\\
I would like to thank Masanori Yamanaka for useful discussions about
related topics. 
This work is supported by Grant-in-Aid for Young Scientists (B)
(18740243), from MEXT, Japan.
%%%%%%%%%%%%%%%%%%%%%%%%%%%%%%%%%%%%%%%%%%%%%%%%%%%%%%%%%%%%%%%%%%%%%%%%%%%%%%%%%%%
\appendix
%%%%%%%%%%%%%%%%%%%%%%%%%%%%%%%%%%%%%%%%%%%%%%%%%%%%%%%%%%%%%%%%%%%%%%%%%%%%%%%%%%%
\section{Appendix}
\label{a:non-vanishing}
In this appendix we shall show that the pairing state $\PhiG$ is 
non-vanishing when the number of holes, $\Nh=|D|+N$, satisfies
$N=|D|-2l_2L_1$ with some integer $0\le l_2\le (L_2-2)/2$.
A similar argument will show that $\PhiG$ is non-vanishing for $2L_1\le N \le |D|$.

It is easy to see that the collection of the states in the
right-hand-side of \eqref{eq:PhiG-expansion} is
orthogonal.
So $\PhiG$ is non-vanishing if one of those terms is non-vanishing.
We shall show that this is the case.
Let 
\begin{equation}
A_1=
 \left\{
  x=(x_1,x_2)~|~1\le x_1 \le L_1, 1\le x_2 \le 2l_2,
                   \mbox{ $x_2$ is odd} 
 \right\}
\end{equation}
and 
\begin{equation}
A_2=
 \left\{
  x=(x_1,x_2)~|~1\le x_1 \le L_1, 1\le x_2 \le 2l_2,
                   \mbox{ $x_2$ is even} 
 \right\}.
\end{equation} 
Now we pick up the state in \eqref{eq:PhiG-expansion} 
corresponding to the subset $S_0=(A_1\cup A_2) \cap \De$.
Substituting $\tphiz=a_{z,\dn}b_{z,\up}+b_{z,\dn}a_{z,\up}$ into this
state,
we obtain
\begin{equation}
 \prod_{z\in S_0}(a_{z,\dn}b_{z,\up}+b_{z,\dn}a_{z,\up})\Psi_0
  =\sum_{T\subset S_0}\left(\prod_{z\in T}a_{z,\dn}b_{z,\up}\right)
                      \left(\prod_{z\in S_0\bs T}b_{z,\dn}a_{z,\up}\right)\Psi_0.
\label{eq:PhiG-expansion2}
\end{equation}
The collection of the states in the right-hand-side of the above expression is again
orthogonal.
Let 
$S_1=A_1\cap \De$.
Then it is easy to see that
\begin{equation}
\left<
 \left(\prod_{z\in A_1}a_{z,\dn}\right)
 \left(\prod_{z\in A_2}a_{z,\up}\right)\Psi_0,
 \left(\prod_{z\in S_1}a_{z,\dn}b_{z,\up}\right)
 \left(\prod_{z\in S_0\bs S_1}a_{z,\up}b_{z,\dn}\right)\Psi_0
\right>
\end{equation}
is non-zero.
This implies that the term in \eqref{eq:PhiG-expansion2} with $T=S_1$ 
(and thus the term with $S=S_0$ in \eqref{eq:PhiG-expansion2}) is
non-vanishing, which concludes that $\PhiG$ is non-vanishing. 
%%%%%%%%%%%%%%%%%%%%%%%%%%%%%%%%%%%%%%%%%%%%%%%%%%%%%%%%%%%%%%%%


\begin{thebibliography}{99}

\bibitem{Bednorz}
	J. G. Bednorz and K. A. M\"uller, Z. Phys. B \textbf{64}, 189 (1986).

\bibitem{Anderson}
	P. W. Anderson, Science \textbf{235}, 1196 (1987).

\bibitem{Emery87}
	V. J. Emery, Phys. Rev. Lett. \textbf{58} 2794 (1987).

\bibitem{Hirsch87}
	J. E. Hirsch, Phys. Rev. Lett. \textbf{59} 228 (1987).

\bibitem{ZhangRice88}
	F. C. Zhang and T. M. Rice, Phys. Rev. B \textbf{37}, 3759 (1988).

\bibitem{Tanaka04}
	A. Tanaka, J. Phys. A: Math. Gen. \textbf{37}, 1559 (2004).

\bibitem{comment1}
	Here $|\cdot|$ represents the Euclidean norm.
	The same symbol $|X|$ is used to denote the number of elements in a
	set $X$.
\end{thebibliography}
\end{document}